\documentclass[useAMS,usenatbib,usegraphicx]{mn2e}

\title[The first spectroscopic L dwarf in Praesepe]{The first
  spectroscopically identified L dwarf in Praesepe}
  \author[S. Boudreault and N.  Lodieu]{S.
  Boudreault$^{1,2}$\thanks{E-mail: szb@iac.es} and N. Lodieu$^{1,2}$\\
  $^{1}$ Instituto de Astrof\'{i}sica de Canarias (IAC), C/V\'{i}a
  L\'{a}ctea s/n, E-38200 La Laguna, Tenerife, Spain\\
  $^{2}$ Departamento de Astrof\'{i}sica, Universidad de La Laguna
  (ULL), E-38205 La Laguna, Tenerife, Spain}

\begin{document}

\date{Accepted 2013 June 5. Received 2013 May 17; in original form 2013 May 1}

\pagerange{\pageref{firstpage}--\pageref{lastpage}} \pubyear{2013}

\maketitle

\label{firstpage}

\begin{abstract}
  We have obtained a low-resolution optical spectrum for one of the
  faintest cluster member candidates in Praesepe with the Optical
  System for Imaging and low Resolution Integrated Spectroscopy
  mounted on the 10.4\,m Gran Telescopio de Canarias. We confirm
  spectroscopically the first L dwarf member in Praesepe, UGCS
  J084510.66$+$214817.1\@.  We derived an optical spectral type of
  L0.3$\pm$0.4 and estimated its effective temperature to $T_{\rm
    eff}$$=$2279$\pm$371\,K and a mass of 71.1$\pm$23.0\,M$_{\rm
    Jup}$, according to state-of-the-art models, placing it at the
  hydrogen-burning boundary. We measured the equivalent width of the
  gravity-sensitive sodium doublet at 8182/8194\AA{}, which adds
  credit to the membership of this new L dwarf to Praesepe. We also
  derived a probability of $\sim$20.5\% that our candidate would be a
  field L0 dwarf. We conclude that this object is likely to be a true
  member of Praesepe, with evidence of being a binary system.
\end{abstract}

\begin{keywords}
open clusters and associations: individual: Praesepe ---
     stars: low-mass, brown dwarfs -- stars: fundamental
     parameters --- techniques: spectroscopic.
\end{keywords}

\section{Introduction}

More than 900 L dwarfs have been confirmed spectroscopically so far,
most of them nearby and
old\footnote{http://spider.ipac.caltech.edu/staff/davy/ARCHIVE/index.shtml,
  as of 6 November 2012\@.} (i.e. $\geq$1\,Gyr).  Nonetheless,
  some of them have been observed in young (i.e.
  $\leq$10\,Myr) star-forming regions and confirmed spectroscopically,
  e.g.\ in Upper Sco \citep{lodieu2008}, Taurus \citep{luhman2008},
  $\sigma$~Ori
  \citep{mcgovern2004,zapatero2000,barrado2001,martin2001},
  Chamaeleon~I \citep{luhman2008}, Trapezium \citep{lucas2006},
  $\rho$\,Oph \citep{oliveira2012}, NGC\,1333 \citep{scholz2012}. To
  the contrary, very few L dwarf candidates have been confirmed
  spectroscopically as L dwarf members of older clusters with ages
  greater than 10\,Myr. Except in the case of the Pleiades where
  near-infrared spectroscopic follow-up of L dwarf candidates has been
  conducted \citep{bihain2010}, no other spectroscopic L dwarf has
  been announced in regions like IC\,2391
  \citep[50\,Myr;][]{barrado2004}, $\alpha$\,Per
  \citep[85\,Myr;][]{barrado2004}, Blanco~1
  \citep[132\,Myr;][]{cargile2010}, or the Hyades
  \citep[625\,Myr;][]{perryman1998}, to cite only a few examples.
  L dwarfs contract and cool with time, becoming fainter with
  age \citep{dantona1985,burrows2001}, making spectroscopic follow-up
  in these old clusters challenging with current
  instrumentation. However, such spectroscopic observations are of
  prime importance to define spectral type vs effective temperature as
  a function of age for L dwarf \citep{kirkpatrick2005,cruz2009} and
  to trace the early evolution of brown dwarfs \citep{luhman2012}.

Praesepe is an important open cluster that can contribute to
  this goal considering its age
  \citep[590$^{+150}_{-120}$\,Myr;][]{fossati2008}, as this would help
  to fill the gap between L dwarfs known in young populations (i.e.
  $\leq$10\,Myr) and the old L dwarfs in the field (i.e.
  $\geq$1\,Gyr).  Praesepe is also in interesting target to search for
  L dwarfs, considering its distance
  \citep[181.97$^{+5.96}_{-5.77}$\,pc;][]{vanleeuwen2009}, significant
  proper motion \citep[$\mu_\alpha\cos{\delta}$\,=\,$-$35.81$\pm$0.29
  mas/yr and $\mu_\delta$\,=\,$-$12.85$\pm$0.24
  mas/yr;][]{vanleeuwen2009}, and the low extinction towards the
  cluster \citep[$E(B-V)$$=$0.027$\pm$0.004\,mag;][]{taylor2006}. In
  \citet{boudreault2012} we presented 1,116 candidate cluster members
  from an analysis of a near-infrared ($ZYJHK$) wide-field survey of
  36 square degrees in Praesepe carried out by the UKIRT Infrared Deep
  Sky Survey \citep[UKIDSS;][]{lawrence2007} Galactic Clusters Survey
  (GCS) and released in the Ninth Data Release (DR9).

We have initiated a spectroscopic follow-up of the homogeneous samples
of low-mass member candidates extracted from the UKIDSS GCS in
Praesepe. The objectives are two-fold (1) obtain well-defined
spectroscopic samples of late-M and early-L dwarfs with known ages and
distances, and (2) confirm spectroscopically the discrepancy observed
between the lower-end of the Praesepe and Hyades mass functions,
namely, a decrease in the Praesepe mass function of only
$\sim$0.2\,dex from 0.6 down to 0.1M$_\odot$, compared to $\sim$1\,dex
for the Hyades \citep{boudreault2012}.

\begin{figure}
  \includegraphics[width=8cm]{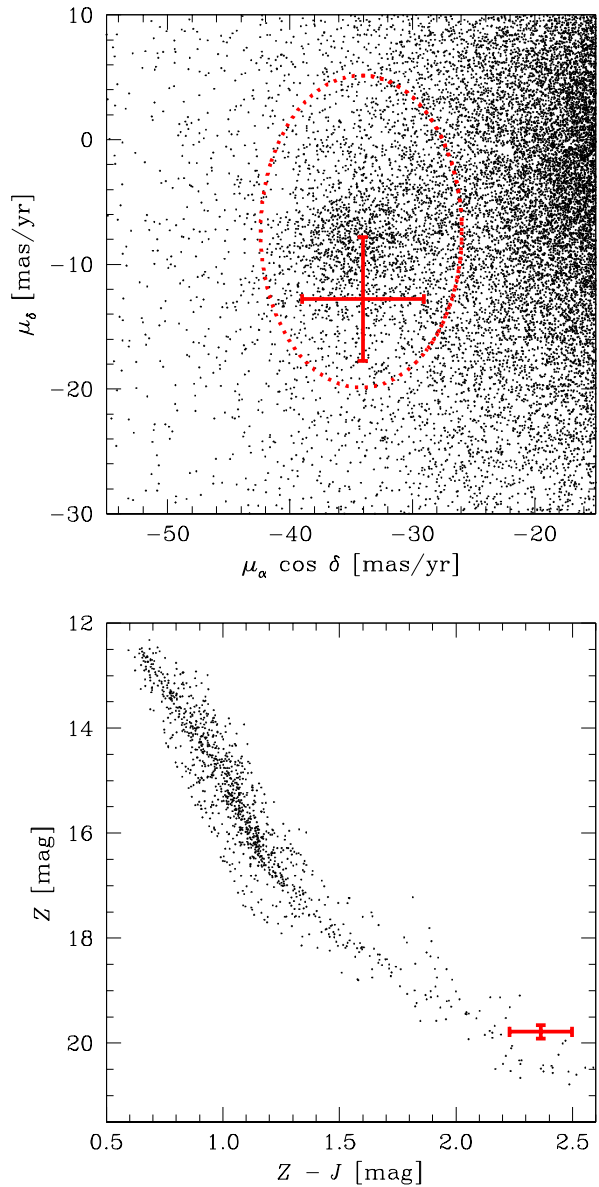}
  \caption{Proper motion diagram (\textit{top panel}) of all
      objects from the UKIDSS GCS DR9 toward Praesepe, and
      colour-magnitude diagram of $Z$ vs. $Z-J$ (\textit{lower panel})
      of all 1,116 Praesepe cluster member candidates based on proper
      motion and photometry \citep{boudreault2012}. The location of
      the L dwarf GCS0845 is indicated by a red cross with its error
      bars on both diagrams. In the proper motion diagram, the red
      ellipse represents 3$\sigma$ \citep[i.e. the astrometric
      selection of][]{boudreault2012} around the location of Praesepe,
      clearly visible at
      $\mu_{\alpha}\cos{\delta}$\,=\,$-$34.17$\pm$2.74 mas/yr and
      $\mu_{\delta}$\,=\,$-$7.36$\pm$4.17 mas/yr.}
  \label{pm-cmd}
\end{figure}

Among our Praesepe candidates, one object has the colours of field L
dwarfs \citep[$Z-J$$=$2.36\,mag and
$J-K$$=$1.39\,mag;][]{hewett2006,schmidt2010}. This object,
UGCS\,J084510.66$+$214817.1 (hereafter GCS0845), is plotted in a
vector-point diagram and a colour-magnitude diagram in Fig.\
\ref{pm-cmd}.

In this letter, we confirm spectroscopically the first L dwarf
selected from our sample of photometric and astrometric cluster member
candidates in Praesepe \citep{boudreault2012}. In Section
\ref{Prae_dL:obs_DR}, we describe the spectroscopy and the data
reduction.  We derive the spectral type in Section \ref{Prae_dL:SpT},
determine the effective temperature ($T_{\rm eff}$) and mass of
GCS0845 in Section \ref{Prae_dL:temp_mass}, and discuss its membership
likelihood in Section \ref{Prae_dL:disc}.

\section{Observations and data reduction}
\label{Prae_dL:obs_DR}

The observations of GCS0845 were carried out in service mode on the
night of 16 January 2013 with the Optical System for Imaging and low
Resolution Integrated Spectroscopy (OSIRIS) mounted on the 10.4\,m
Gran Telescopio Canarias (GTC) in the Roque de Los Muchachos
Observatory in La Palma (Canary Islands). We used the R300R grism and
a 1.0\,arcsec slit with a 2$\times$2 binning, yielding a spectral
resolution of R\,=\,348 at 6865\AA{}. This configuration shows
contamination from the second order light redwards of 9000\AA{}.
Therefore, we restrain the range of analysis of our spectra to the
6000--9000\AA{} wavelength range\footnote{The signal of GCS0845 is too
  weak below 6000\AA{} and all five spectra taken from the literature
  are available only longwards of $\sim$6300\AA{}.}.  We observed our
target with six on-source integrations of 700\,sec on the same night
with a dither pattern of 10\,arcsec along the slit, giving a total
exposure of 1h10m.

\begin{figure}
  \includegraphics[width=8cm]{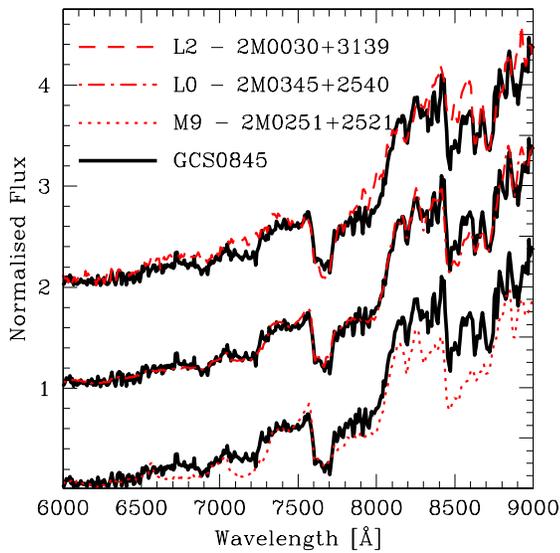}
  \caption{\label{spectra-field} GTC/OSIRIS optical spectrum of
    GCS0845 (black thick line), classified as a L dwarf and a cluster
    member candidate of Praesepe, based on proper motion and
    photometry. Overplotted are the spectra of field dwarfs
    \citep{kirkpatrick1999} observed with the same instrumental
    configuration: 2M0251$+$2521 (M9; red dotted line), 2M0345$+$2540
    (L0; dash-dotted line), and 2M0030$+$3139 (L2; red dash line).
    All spectra are normalised at 7500\AA\ with offset of $+1$ between
    each spectra for clarity.}
\end{figure}

We also observed three field dwarfs with known optical spectral types
with the same instrumental setup: 2M0251$+$2521 (M9), 2M0345$+$2540
(L0), and 2M0030$+$3139 \citep[L2;][]{kirkpatrick1999}. They are
overplotted on the spectrum of our target as red lines in
Fig~\ref{spectra-field}. We observed field L dwarfs with
  spectral types bracketing our target to derive a spectral type. Also
  we can compare the EW of the gravity-sensitive Na{\small{I}} doublet
  in our cluster candidate with the ones of field dwarfs to better
  constrain the membership of GCS0845. Log of the observations are
  provided in Table \ref{tab:info-obs}.

The spectra were reduced using standard IRAF packages
\citep{tody1986,tody1993}. First, the standard CCD reduction steps
(overscan subtraction, trimming, and flat-fielding) were performed on
a nightly basis using the \textit{ccdred} package. After extraction of
the spectra, we performed the wavelength calibration using a
combination of HgAr, Ne and Xe lamps. We calibrate our targets in flux
with the spectrophotometric standard star G191--B2B \citep{oke1990}
and Ross\,640 \citep{oke1974}.

\begin{table}
  \begin{minipage}[t]{\columnwidth}
    \caption{Dates of observations, number of exposures (nbr.),
      exposure times, and signal-to-noise ratio (SNR) of the spectra
      at $\sim\lambda\lambda$8600\AA{}.}
    \label{tab:info-obs}
    \centering
    \renewcommand{\footnoterule}{} 
    \begin{tabular}{@{\hspace{0mm}}l @{\hspace{2mm}}c @{\hspace{2mm}}c @{\hspace{2mm}}c @{\hspace{0mm}}}
      \hline
      Object & Date of observations & Exp. time & SNR \\
      & dd/mm/yyyy & nbr. $\times$ sec &  \\
      \hline
      GCS0845 & 16/01/2013 & 6$\times$700\,sec & 34 \\
      G191--B2B & 16/01/2013 & 1$\times$7.5\,sec & $>$120 \\
      \hline
      2M0251$+$2521 (M9) & 22/08/2012 & 1$\times$60\,sec & 41 \\
      2M0345$+$2540 (L0) & 22/08/2012 & 1$\times$60\,sec & 36 \\
      2M0030$+$3139 (L2) & 22/08/2012 & 1$\times$60\,sec & 63 \\
      Ross 640 & 22/08/2012 & 1$\times$60\,sec & $>$120 \\
      \hline
    \end{tabular}
  \end{minipage}
\end{table}

\section{Optical spectral type}
\label{Prae_dL:SpT}

\begin{figure}
  \includegraphics[width=8cm]{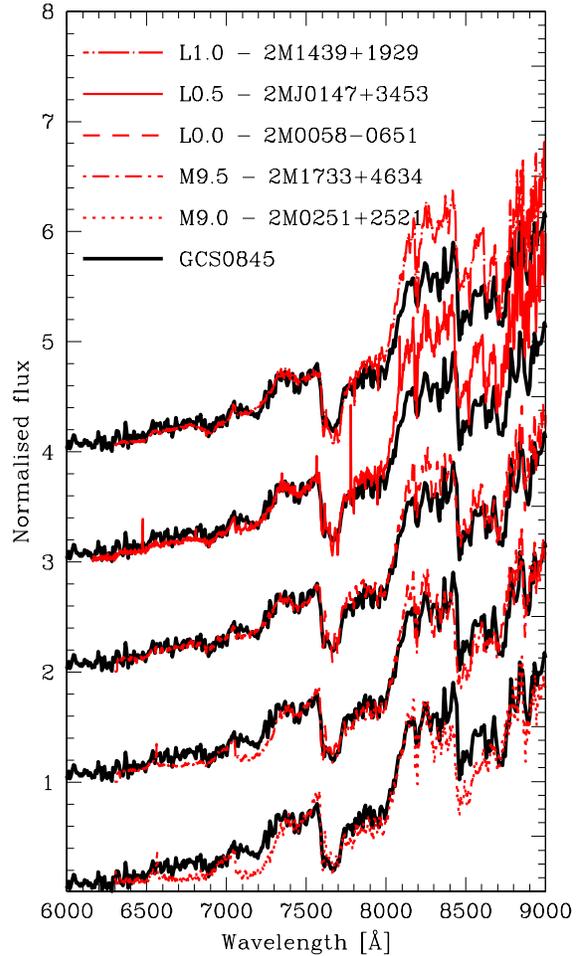}
  \caption{Same as Fig.\ \ref{spectra-field}, but the spectrum of
    GCS0845 is compared to those of other field dwarfs with known
    spectral types taken from the literature not observed with
    GTC/OSIRIS: the 2M0251$+$2521 \citep[M9.0;][]{kirkpatrick1999},
    2M1733$+$4634 \citep[M9.5;][]{gizis2000}, 2M0058$-$0651
    \citep[L0.0;][]{kirkpatrick2000}, 2MJ0147$+$3453
    \citep[L0.5;][]{kirkpatrick1999}, and 2M1439+1929
    \citep[L1.0;][]{kirkpatrick1999} from bottom to top. All
      spectra are normalised at 7500\AA\ with offset of $+1$ between
      each spectra for clarity.}
  \label{spectra-field-other}
\end{figure}

We derived the spectral type of GCS0845 in three ways. First, we
directly compared the optical spectrum of our target with field dwarfs
with known spectral types (Fig.~\ref{spectra-field} and
\ref{spectra-field-other}).  Second, we computed the root-mean-square
(RMS or residuals) between the spectrum of GCS0845 and those of field
dwarfs with known spectral types, normalising them at different
wavelengths (Table \ref{tab:rms-field}). Third, we computed spectral
indices (Table \ref{tab:spectral-indices}) defined in the literature
\citep{kirkpatrick1999,martin1999}. The field dwarfs with known
spectral types used in our analysis are all taken form the literature
(see below for references).

In Fig.\,\ref{spectra-field}, we present the spectrum of GCS0845 and
the spectra of the field dwarfs \citep{kirkpatrick1999} observed with
OSIRIS: 2M0251$+$2521 (M9), 2M0345$+$2540 (L0), and 2M0030$+$3139
(L2). Visually, the best agreement is obtained for the L0 dwarf rather
than for the M9 or L2 dwarfs. We also compared GCS0845 with other
field dwarfs of known spectral types from the literature
(Fig.\,\ref{spectra-field-other}): 2M0251$+$2521
\citep[M9.0;][]{kirkpatrick1999}, 2M1733$+$4634
\citep[M9.5;][]{gizis2000}, 2M0058$-$0651
\citep[L0.0;][]{kirkpatrick2000}, 2MJ0147$+$3453
\citep[L0.5;][]{kirkpatrick1999}, and 2M1439+1929
\citep[L1.0;][]{kirkpatrick1999}. Again, the best match is for a L0
type. In both Figures \ref{spectra-field} and
  \ref{spectra-field-other}, all spectra are normalised at 7500\AA.
  This value was chosen due to the small amount of absorption at this
  wavelength \citep{kirkpatrick1991}.

We computed the residuals between the spectrum of our target and those
of the field dwarfs observed with OSIRIS\@. We used a normalisation in
six different wavelength regions with a width of 25\AA{} and centered
at 6500\AA{}, 7000\AA{}, 7500\AA{}, 7900\AA{} and 8250\AA{}. This
allows us to (1) know which reference star gives the best
agreement at different normalised regions, and (2) to confirm,
  quantitatively, that the spectral type determination is not
  influenced by the normalisation point. The RMS values are presented
  in Table \ref{tab:rms-field} both for the field dwarfs observed with
  the same instrumental configuration and field dwarfs from the
  literature. We observe that the best agreement with our optical
  spectrum is for the L0 dwarf 2M0345$+$2540 normalised at 6500\AA{},
  7900\AA{} and at 8250\AA{}. On the other hand, the best agreement
  with the spectra taken from the literature is with the L1 dwarf
  2M1439+1929 normalised at 7500\AA{}. However, considering the other
  similar RMS values for the L0 dwarf 2M0058$-$0651 with normalisation
  at 7500\AA{} and for the L0.5 dwarf 2M0147$+$3453 with normalisation
  at 8250\AA{}, we consider that the agreement taken with spectra from
  the literature span from L0 to L1, hence assuming an L0.5.

\begin{table*}
  \begin{minipage}[t]{\linewidth}
    \caption{RMS values for various regions of normalisation (column
      1) between GCS0845 and the field dwarf presented in
      Fig.~\ref{spectra-field}.  We also give the RMS values for the
      field dwarfs presented in Fig.~\ref{spectra-field-other}.}
    \label{tab:rms-field}
    \centering
    \renewcommand{\footnoterule}{} 
    \begin{tabular}{@{\hspace{2mm}}c| @{\hspace{2mm}}c @{\hspace{2mm}}c @{\hspace{2mm}}c| @{\hspace{2mm}}c @{\hspace{2mm}}c @{\hspace{2mm}}c @{\hspace{2mm}}c @{\hspace{2mm}}c @{\hspace{0mm}}}
      \hline
      $\lambda$ & 0251$+$2521 & 0345$+$2540 & 0030$+$3139 &
      0251$+$2521 & 1733$+$4634 & 0058$-$0651 & 0147$+$3453 & 1439+1929 \\
      & (M9) & (L0) & (L2) & (M9) & (M9.5) & (L0) & (L0.5) & (L1) \\ 
      \hline
      6500\AA &  0.67 &  0.34 &  0.42 &  0.82 &  0.79 &  0.66 &  1.16 &  0.75 \\
      7000\AA &  0.52 &  0.40 &  0.57 &  0.87 &  0.65 &  0.51 &  0.99 &  0.73 \\
      7500\AA &  1.02 &  0.41 &  0.48 &  1.16 &  0.88 &  0.44 &  0.56 &  0.42 \\
      7900\AA &  0.49 &  0.34 &  0.48 &  0.60 &  0.56 &  0.62 &  0.45 &  0.49 \\
      8250\AA &  0.48 &  0.34 &  0.47 &  0.63 &  0.59 &  0.46 &  0.45 &  0.45 \\
      \hline
    \end{tabular}
  \end{minipage}
\end{table*}

Finally, we used three spectral indices (CrH--a, Rb--b/TiO--b, and
Cs-a/VO--d) from \citet{kirkpatrick1999} defined in their Table 7 and
also the PC3 spectral index from \citet{martin1999}.  We report the
values of the spectral indices and their associated spectral types in
Table \ref{tab:spectral-indices}. The spectral indices CrH--a and
Rb--b/TiO--b are consistent with the ones of a L0 and a L2 dwarf
respectively, while the spectral index Cs--a/VO--b suggests a later
type closer to L1\@.  The PC3 spectral index suggests a spectral type
of L0.1 or L1.4\@.

\begin{table*}
  \begin{minipage}[t]{\linewidth}
    \caption{Values of spectral indices \citep[CrH--a, Rb--b/TiO--b,
      Cs--a/VO--b;][]{kirkpatrick1999} for our Praesepe member, field
      dwarfs observed with OSIRIS, and field dwarfs from the
      literature. We also list the values for the PC3 spectral index
      \citep{martin1999} using both expressions for spectral types
      between M2.5 and L1 (left values) and for spectral types between
      L1 and L6 (right values). This is done for our target and all
      objects with spectral type within L0 and L2. For the M9 and M9.5
      field dwarfs, we only use the first equation of
      \citet{martin1999}. We give in brackets the expected spectral
      types for all objects, based on each value of the spectral
      indices. For the spectral indices from \citet{kirkpatrick1999},
      we estimate the spectral type based on values presented in their
      Table 8A\@. The last column gives the average (Avg) of the
      spectral types derived from all indices. We can see that all
      average spectral type match within $\sim$0.5 subclass those
      expected (left column).}
    \label{tab:spectral-indices}
      \centering
      \renewcommand{\footnoterule}{} 
    \begin{tabular}{l|cccc|c}
      \hline
      Object & CrH--a & Rb--b/TiO--b & Cs--a/VO--b & PC3 & Avg \\
      \hline
      GCS0845 & 1.254[L0] & 0.984[L2] & 0.840[L1] & 10.144[L0.1] or 11.355[L1.4] & L0.9 \\
      \hline
      2M0251$+$2521 (M9) & 1.010[M7] & 0.590[M9] & 0.711[M9] & 8.328[M8.3] & M8.3 \\
      2M0345$+$2540 (L0) & 1.118[L0] & 0.656[L0] & 0.708[M9] & 9.935[M9.9] or 11.206[L1.2] & L0.0 \\
      2M0030$+$3139 (L2) & 1.402[L1] & 0.916[L2] & 0.970[L2] & 10.26[L0.3] or 11.532[L1.5] & L1.4 \\
      \hline
      2M0251$+$2521 (M9) & 0.978[M7] & 0.694[L0] & 0.774[M9] & 8.473[M8.5] & M8.6 \\
      2M1733$+$4634 (M9.5) & 1.096[M9] & 0.670[L0] & 0.743[M9] & 9.478[M9.5] & M9.4 \\
      2M0058$-$0651 (L0) & 1.164[L0] & 0.783[L1] & 0.747[M9] & 10.066[L0.1] or 11.290[L1.3] & L0.3 \\
      2M0147$+$3453 (L0.5) & 1.278[L1] & 0.800[L1] & 0.788[L0] & 10.266[L0.3] or 11.571[L1.6] & L0.8 \\
      2M1439$+$1929 (L1) & 1.365[L1] & 0.883[L1] & 0.845[L1] & 10.211[L0.2] or 11.730[L1.7] & L1.2 \\
      \hline
    \end{tabular}
  \end{minipage}
\end{table*}

We took the average of all the spectral type determinations
(2$\times$L0 from the optical visualisation, L0 and L0.5 from the
residuals, and L0.9 from the spectral indices) and used the standard
deviation to estimation of the error. We derive a spectral type of
L0.3 with a standard deviation of 0.4 subclass for GCS0845\@.

\section{Effective temperature and mass}
\label{Prae_dL:temp_mass}

We used the average and standard deviation of our spectral
  type (L0.3$\pm$0.4) to derive the spectroscopic $T_{\rm eff}$ of
  GCS0845\@.  We used three $T_{\rm eff}$ vs spectral type relations
  to derive the effective temperature: from \citet{stephens2001},
  \citet{burgasser2001}, and \citet{dahn2002}.  The $T_{\rm eff}$ vs
  spectral type relation of field L dwarfs from \citet{stephens2001}
  gives 2190$\pm$40\,K, while the $T_{\rm eff}$ vs spectral type
  relation from \citet{burgasser2001} gives us 2338$\pm$98\,K.  No
  equations of $T_{\rm eff}$ vs spectral type are provided by
  \citet{dahn2002}. Therefore, we apply a linear regression between
  the $T_{\rm eff}$ and spectral type values of the M and L dwarfs
  with known parallaxes from \citet[][; their Table 5]{dahn2002}. With
  this we obtain $T_{\rm eff}$\,=\,2310$\pm$356\footnote{The
      error on this value of $T_{\rm eff}$ is higher because, in
      addition to the error on the spectral type, we include the error
      on our linear regression of the values from
      \citet{dahn2002}.}\,K. Taking the average of these estimates
  and adding the errors in quadrature, we derive $T_{\rm
    eff}$\,=\,2279$\pm$371\,K for GCS0845\@.

With the derived $T_{\rm eff}$ and using an evolutionary track for
objects with dusty atmospheres \citep[DUSTY;][]{chabrier2000} at the
age of Praesepe \citep[590$^{+150}_{-120}$\,Myr;][]{fossati2008}
following \citet{boudreault2012}, we inferred a spectroscopic mass of
71.6$\pm$15.0\,M$_{\rm Jup}$ for GCS0845\@. If we repeat the same
procedure with the recent BT-Settl models \citep{allard2012}, we
obtain a mass of 70.5$\pm$17.5\,M$_{\rm Jup}$. The errors on the
masses take into account the uncertainties on the age of the cluster.
Taking the average of these values, we derive a mass of
71.1$\pm$23.0\,M$_{\rm Jup}$, which places GCS0845 at the
stellar/substellar boundary at $\sim$70\,M$_{\rm Jup}$
  \citep[assuming a solar composition;][]{chabrier2000araa}.
  The lithium feature at 6708\AA{} would address its
  substellarity. However, the low signal-to-noise ratio at the
  location of Li{\rmfamily\scshape I} at $\lambda\lambda$6708\AA{}
  does not allow us to draw conclusions about its presence. This, and
  considering the error on the mass determination of GCS0845, it not
  possible to constain its stellar or substellar nature.

\section{Discussion on membership}
\label{Prae_dL:disc}

As mentioned previously, the low signal-to-noise ratio at the
  location of the H$\alpha$ at $\lambda\lambda$6563\AA{} does not
  allow us to draw conclusions about its presence or to measure its
  equivalent width (EW). Because of our low spectral
  resolution, we do not resolve the gravity-sensitive Na{\small{I}}
  doublet at 8182/8194\AA{} because of the low spectral resolution of
  our spectra.  We measured an EW of 6.0$\pm$2.5\AA{} for the
  unresolved Na{\small{I}} doublet of GCS0845 whereas we obtained
  11.4$\pm$0.5\AA{}, 8.4$\pm$1.2\AA{}, and 18.4$\pm$1.8\AA{} for the
  field dwarfs 2M0251$+$2521, 2M0345$+$2540 and 2M0030$+$3139. The
  EW of the gravity-sensitive feature of GCS0845 is 1$\sigma$
  lower than the field L0 dwarf 2M0345$+$2540, more than 2$\sigma$
  lower than the field M9 dwarf 2M0251$+$2521, and more than 4$\sigma$
  lower than field L2 dwarf 2M0030$+$3139. We advocate cautiousness
  with these measurements, considering that the Na{\small{I}} doublet
  is not resolved but only its merged feature at
  8188\AA. However, our
  measurements point towards an object with a lower surface gravity
  than the field objects. This is consistent with the age of Praesepe
  (590$^{+150}_{-120}$\,Myr) being younger than those of the field
  dwarfs (i.e.  $>1$\,Gyr)\footnote{Based on the BT-Settl
    models \citep{allard2012}, an object of 590\,Myr in Praesepe with
    $T_{\rm eff}$\,$\sim$\,2300\,K would have a surface gravity
    $\sim$0.1\,dex lower than a 1\,Gyr field object with the same
    $T_{\rm eff}$.}.

In order to establish a membership probability for GCS0845, we used
the local density of field L0 dwarfs around the Sun from
\citet{caballero2008}, which is $\sim$7$\times$10$^{-4}$ per pc$^3$.
We defined a total volume of $\sim$23,450\,pc$^3$ for our survey of
Praesepe, considering its coverage \citep[$\sim$36 square
degrees;][]{boudreault2012} and that the cluster candidates span
$\sim$1\,mag at around the $Z-J$ colours of GCS0845
(Fig.\,\ref{pm-cmd}). We find $\sim$16.4 L0 field dwarfs at the
location of Praesepe, which would have been mistakenly considered as
cluster members with photometry only. On the other hand, a field L0
dwarf at the location of Praesepe could have a random proper motion as
high as 67\,mas/yr \citep[corresponding to a tangential velocity of
$\sim$100\,km/s, which is observed for some late-type dwarfs in the
field;][]{faherty2012}, giving a total coverage of
$\sim$14,000\,mas/yr$^2$ in the vector point diagram, while our target
is within 1.5$\sigma$ around the mean motion of Praesepe, yielding
175\,mas/yr$^2$ (i.e.  1.25\%). Hence, we estimate a total of
$\sim$0.205 L0 field dwarf with photometry and astrometry consistent
with Praesepe, equivalent to a probability of $\sim$20.5\% that our
target is a field dwarf. This number is consistent with the upper
limit of contamination in the astrometric and photometric selection of
cluster candidate members from \citet{boudreault2012}: 23.8\% for
masses below 0.15\,M$_\odot$.

We used the calibrated absolute magnitude $M_J$ vs spectral type
relations from \citet{hawley2002} and \citet{dahn2002} to derive the
observed magnitude of GCS0845 at the distance of Praesepe (m-M\,=\,6.3
mag), obtaining $J$\,=\,18.15$\pm$0.15 mag and $J$\,=\,18.23$\pm$0.17
mag, respectively.  The errors on the $J$ magnitudes reflect the
uncertainty on our spectral type determination. These values are not
consistent with the observed magnitude of GCS0845 in UKIDSS GCS
($J$\,=\,17.42$\pm$0.03 mag).  However, these numbers would match with
the observed magnitude of GCS0845 if this object is a binary
system. The later would be consistent with the location of GCS0845 in
the ($Z-J$,$Z$) colour-magnitude diagram in Fig.\ \ref{pm-cmd}: the
cluster sequence is observed from ($Z-J$,$Z$)\,$\sim$\,(1.3,17.0) to
(2.3,20.5) with the binary sequence $\sim$0.75 mag above it, where
GCS0845 lies.

Considering that GCS0845 was originally selected as a photometric and
astrometric member \citep{boudreault2012}, has a spectral type
consistent with the colours of a L dwarf, has a gravity-sensitive
feature suggesting an age younger than $\sim$1\,Gyr, and has a
$\sim$79.5\% probability of belonging to Praesepe, we conclude that
GCS0845 is most likely a member of Praesepe.  However, absolute
magnitude $M_J$ vs spectral type relations do not support this
conclusion, unless GCS0845 is a binary system, an argument also
suggested by its location in the ($Z-J$,$Z$) diagram.

GCS0845 represents a new L dwarf with a known age and distance that is
amenable to further investigation. This object will also guide current
and future deep pencil-beam surveys to look for bona-fide brown dwarf
members in Praesepe.

\section*{Acknowledgments}

SB and NL are funded by national program AYA2010-19136 (Principal
Investigator is NL) funded by the Spanish ministry of science and
innovation. NL is a Ram\'on y Cajal fellow at the IAC (program
08-303-01-02).  We are grateful to Antonio Cabrera-Lavers for keeping
us up-to-date with the GTC observations.  This research has made use
of the Simbad database, operated at the Centre de Donn\'ees
Astronomiques de Strasbourg (CDS), and of NASA's Astrophysics Data
System Bibliographic Services (ADS).

\bibliographystyle{mn2e}
\bibliography{./mnemonic,./boudreault-ldwarf-praesepe}

\label{lastpage}

\end{document}